# Periodic optical rogue waves

Yingchun Ding,[1] Junbo Gao,[1] Fengli Zhang,[1] Zhaoyang Chen[*],[1] Chengyou Lin[1] and M. Y. Yu[2,3]

[1]Department of Physics, Beijing University of Chemical Technology, Beijing, 100029, China.
[2]Institute for Fusion Theory and Simulation, Zhejiang University, Hangzhou 310027, China
[3]Institute for Theoretical Physics I, Ruhr University, D-44780 Bochum, Germany

**Abstract**

Random excitation of intense periodic highly-localized single-cycle light pulses in a stochastic background by continuous-wave stimulated Brillouin scattering in long optical fibers with weak feedback is found experimentally. Events with low period numbers are dominant and the optical feedback is crucial for the phenomenon. A three-wave coupling model for the phenomenon is proposed. The results are in good qualitative agreement with the observed phenomenon. The latter should be relevant to the understanding of similar rogue wave events in other nonlinear dissipative systems.

---

[*] Corresponding author. Email address: chenzy@mail.buct.edu.cn

***Introduction*** - Optical rogue waves (ORWs) analogous to the ocean rogue waves [1,2] were first found in highly nonlinear optical fibers by Solli et al. [3]. Since then ORWs have received much attention [4,5,6] and experiments on ORWs in doped fibers [7,8], nonlinear optical cavities [9], fiber Raman amplifiers [10], as well as linear systems [11], have been carried out. Proposed mechanisms for the ORWs include modulational instabilities [12-14], soliton collisions [15-18], optical wave turbulence [19,20], hypercycle amplification [9], etc. These mechanisms are most frequently described by various forms of the nonlinear Schrödinger equation (NLSE) [21,22] for modulation and evolution of nonlinear wave envelopes. Other descriptions include the rate equations [23] for excitation and interaction of waves, three-wave coupling equations for Raman, Brilluion, and other scattering [10,24], etc.

Stimulated Brillouin scattering (SBS) is the result of nonlinear excitation of a Stokes sideband and an acoustic wave by an intense light wave [24]. Because of its low threshold and high gain, SBS is one of the most important effects in optical media [25,26]. In this Letter, we report the first experimental evidence of periodic solitary optical pulses that occur stochastically. When the pump power sufficiently exceeds the SBS threshold, the occurrence probability of such events can be quite high. An analytical three-wave coupling model including weak pump feedback is proposed for the observed phenomenon. The numerical results from our model are in good agreement with that of the experiments.

***Experiment*** - Our experimental setup is illustrated in Fig. 1(a). The pump source is a tunable semiconductor laser delivering a partially coherent quasi-continuous wave (CW) output with spectral line width ~50 KHz and minimum output power 10 mW. A polarization-insensitive optical isolator (ISO) that ensures unidirectional laser emission at 1550 nm and an erbium-doped fiber amplifier (EDFA) for increasing the average laser power $P_0$ to ~200 mW are used. The pump light is injected into the fiber via a three-port optical circulator, which also delivers the output from the fiber to the detection system. A conventional single-mode fiber (SMF) of total length $L = 10 \text{ km}$, loss coefficient $\alpha \approx 0.077 \text{ km}^{-1}$, and refractive index $n = 1.46$ is used. One end of the fiber is cleaved at 90°, so that there is a 4% optical pump feedback. The detection system includes an optical attenuator (ATT), a photodiode (PD), and an optical sampling oscilloscope (OSO). After being attenuated by the ATT, the output signal is probed by the PD and temporally characterized by the OSO, so that direct real-time measurement of the intensity profile can be made.

We investigate the nonlinear response of the optical fiber to laser light slightly and well above the SBS threshold power $P_{th} \left( = 15 \text{ mW} \right)$. A typical oscilloscope trace spans

500 $\mu s$. A large set of data (about 6000 waveforms of the output signals) for input pump powers from 20.53 to 68.78 mW have been collected. Fig. 1(b) shows a normal event of SBS in a chaotic background for $P_0 = 15.81$ mW, slightly above the threshold. Fig. 1(c) for $P_0 = 25.43$ mW shows an event with periodic single-cycle pulses in a stochastic background, like soliton chains [28,29]. One can see five evenly spaced sharp single-cycle pulses, each of the same short duration. Their intensities are well above the average as well as that of the normal SBS event shown in Fig. 1(b). A histogram of the probability distribution for the occurrence of such events is given in Fig. 1(c), obtained from a statistical analysis of our entire experimental data set. The occurrence probability increases with $P_0$ up to a maximum of 27.8% at $P_0 = 51.94$ mW, thereafter it first decreases slightly with $P_0$ and then remains roughly constant within the power range investigated.

Since they are of very large amplitude and their occurrence is unpredictable and space-time localized, the intense short-pulse chains observed here may be considered as periodic optical rogue waves (PORWs). However, their occurrence probability with respect to the input pump power is neither gaussian nor L-shaped [27]. In fact, it is gaussian for small pump powers and nearly flat (i.e., partial L-shape) for large pump powers. In the mid-power range its profile is rather complex and involves at least two different scalings. This makes the PORWs somewhat different from the classical ORWs [3-20], whose occurrence probability has usually an L-shaped distribution.

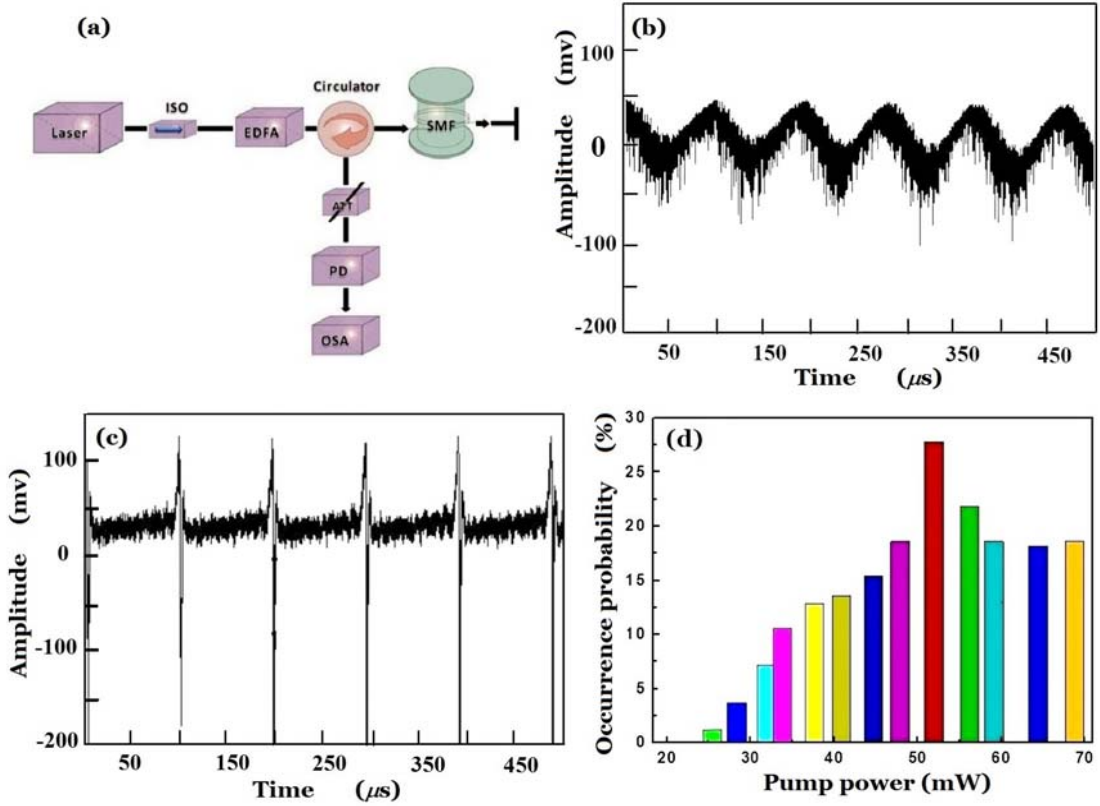

FIG. 1. (color online) (a) Experimental setup (see text for details). (b) Oscilloscope trace of SBS for $P_0 = 15.81$ mW, slightly above the SBS threshold power $P_{th}\ (=15\ \text{mW})$. Here only normal SBS in a high-frequency chaotic background can appear. (c) Oscilloscope trace for $P_0 = 25.43\ \text{mW} \gg P_{th}$. One can see here a PORW event with periodic spikes in a chaotic background. (d) Probability distribution of the PORW events for $25.43 < P_0\ [\text{mW}] < 70$.

Fig. 2 shows PORW events for $P_0 = 25.43$ to $68.78$ mW. We see that higher input pump powers lead to more periodic pulses within the same sweep time interval. Fig. 2(d) gives the occurrence probability distribution of the PORW events for fixed $P_0\ (=68.78\ \text{mW})$. One sees that the 5 peak case shown in Fig. 1(c) is dominant.

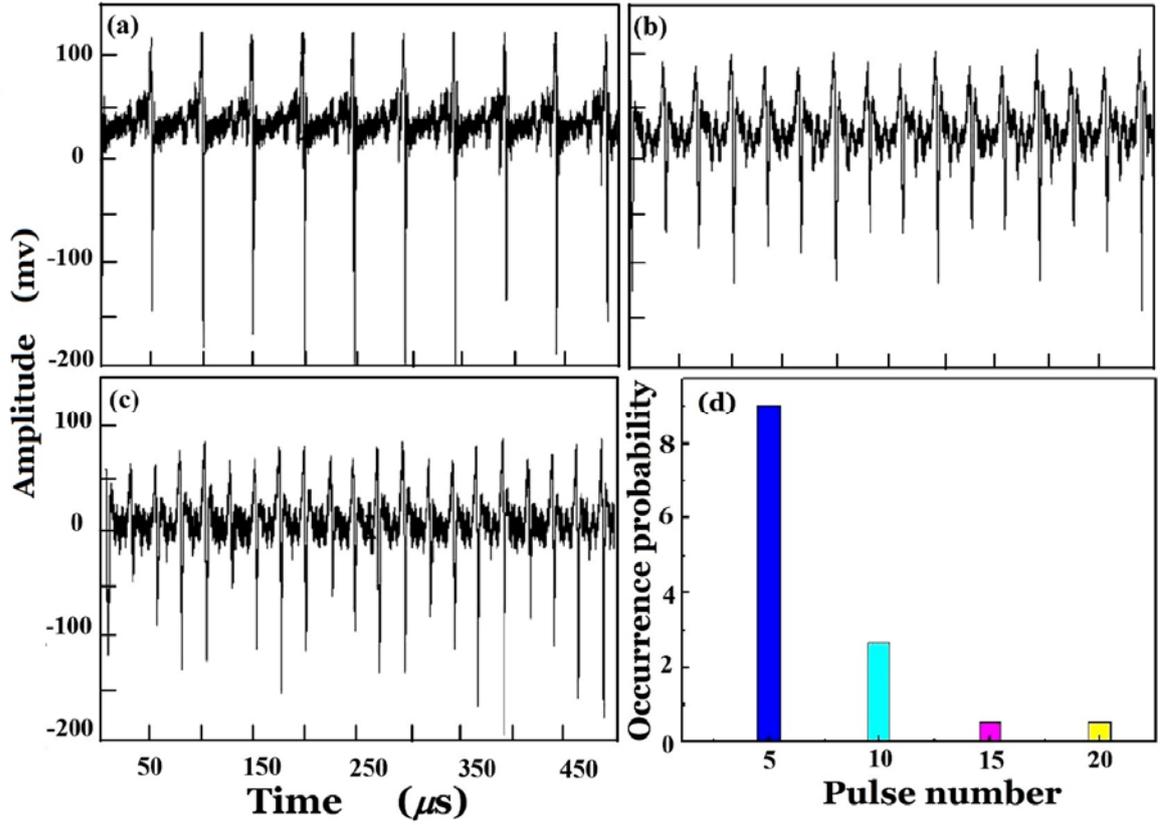

Fig. 2. (Color online) PORWs for $P_0$ [mW]= 33.82 (a), 44.74 (b), and 68.78 (c). The numbers of pulses in the given time interval are 5, 10, 15 and 20, respectively. (d) Histogram of the statistical distribution (in %) of PORW pulses for $P_0$=68.78 mW.

*Analytical model* - We now introduce an analytical model for the PORWs. It is based on the well-known three-wave coupling equations often used for studying nonlinear wave interaction and scattering [30,31]. Taking into account the interaction between forward pump, backward Stokes and acoustic wave, the model equations can be written as

$$\frac{\partial A}{\partial \eta}+\frac{\partial A}{\partial \xi}-iu\left(|A|^2+2|B|^2\right)A+\frac{1}{2}\beta A=-gBC-\delta g|A|^2, \quad (1)$$

$$\frac{\partial B}{\partial \eta}-\frac{\partial B}{\partial \xi}-iu\left(|B|^2+2|A|^2\right)B+\frac{1}{2}\beta B = gAC^*, \quad (2)$$

$$\frac{1}{\beta_A}\frac{\partial C}{\partial \eta}+C = AB^*, \quad (3)$$

where *A* and *B* are the normalized (by the pump amplitude $E_0$ at $z=0$) envelopes of the pump and Stokes waves, and *C* is normalized amplitude of the acoustic wave.

Furthermore, $\eta = t/T_r$ and $\xi = z/L$ are normalized time and space coordinates, $L$ is the interaction length, and $T_r = nL/c$ is the transit time of the light waves in the fiber. $\beta = \alpha L$ and $\beta_A = \pi \Delta \nu_B T_r$ are the normalized power loss of light waves and the relaxation rate of the sound waves, $\alpha$ is the loss coefficient, $\Delta \nu_B$ is the spontaneous Brillouin line width, $g$ is the normalized small-signal gain amplitude of the stimulated scattering, and $u \equiv n_2 \omega L |E_0|^2$, where $n_2$ is the nonlinear refractive index coefficient. The phase modulation parameter $u_0$ is determined by $u = gu_0$. In Eq. (1), we have also included the term $-\delta g A A^*$ for the phase controlled pump feedback, where $\delta$ is the feedback coefficient. This pump feedback plays an important role in the excitation of the PORWs. In fact, when the pump feedback was removed in the experiment, we got only noise.

Equations (1) – (3) are solved using a finite-difference scheme and the experimental parameters $n = 1.46$, $L = 10$ km, $\alpha = 0.077$ km$^{-1}$, $\beta = 0.77$, $\lambda_p = 1550$ nm, $\Delta \nu_B = 50$ MHZ, $\beta_A = 7645.9$, and $g = 1.7 \times 10^3 P_0$. Figs. 3 (a)–(d) show the theoretical results for input pump powers $P_0 = 20, 30, 40$ and $70$ mW, respectively. As in the experiments, here we can clearly see the PORWs in a continuous but stochastic background. The number of pulses in the $500$ $\mu$s time window also increases with $P_0$, also in agreement with the experimental results. However, the amplitude of the PORW pulses obtained from the model are even more sharply peaked than that observed in the experiments. Fig. 3(e) for $\delta = 0.5$ shows only stochastic noise. However, a PORW event occurs when a slightly larger $\delta$ $(=0.6)$ is used, as shown in Fig. 3(f). The crucial role of the pump feedback is thus confirmed. In fact, we see that not only the feedback is necessary, but there is also a threshold value of $\delta$, for the appearance of PORWs

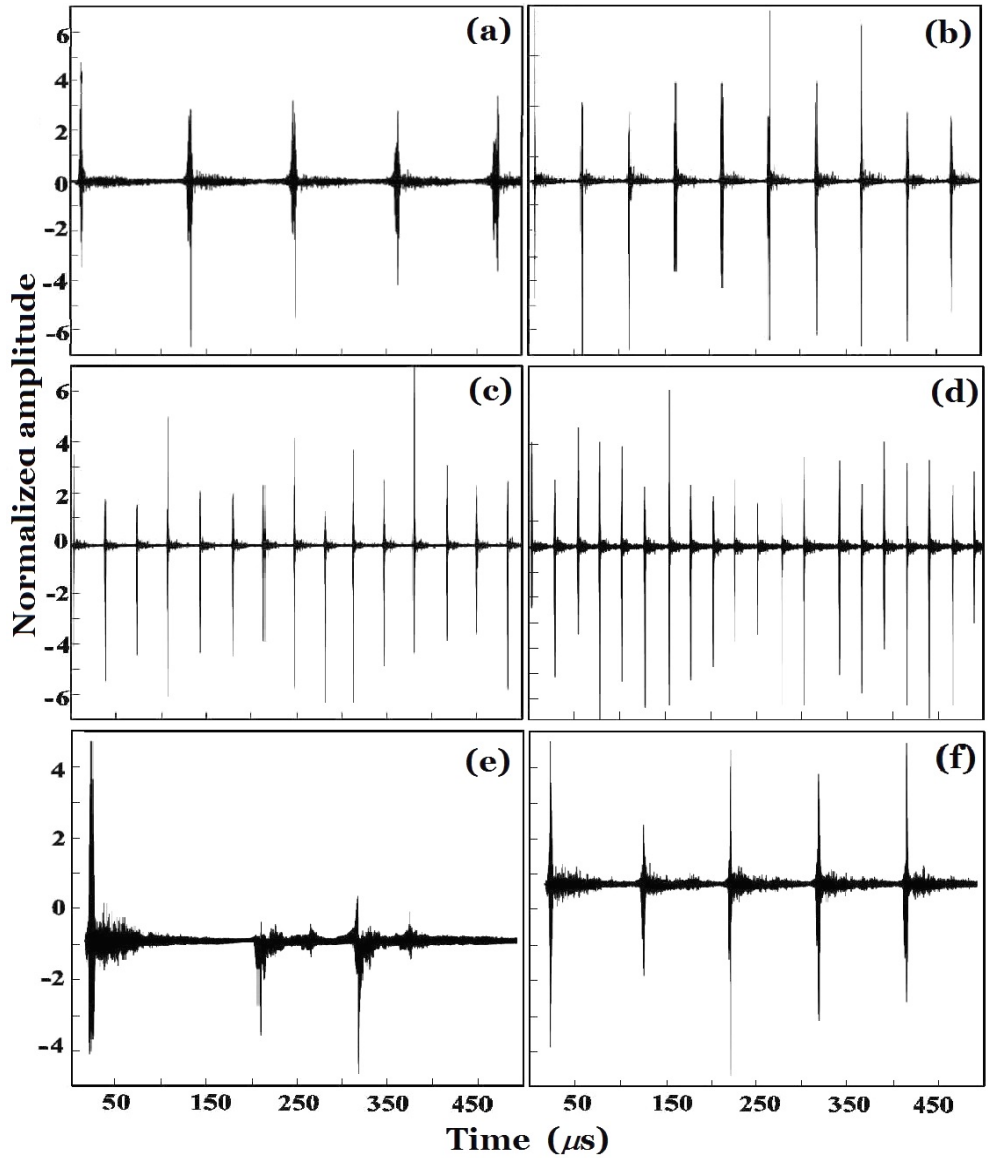

FIG. 3. Solutions of the three-wave coupling model showing the PORWs for $P_0$ [mW] = 20 (a), 30 (b), 40 (c), and 70 (d), $\delta = 0.55$ and $u = 0.7$. The slightly modulated sharp pulses observed in the experiments are recovered by the model, but they are even narrower. The behavior of the weak stochastic background is also slightly different. The panels (e) and (f) show the time series for $\delta = 0.5$ and $\delta = 0.6$, respectively, for the phase modulation parameter $u = 0.7$ and $P_0 = 22$ mW, confirming the importance of optical feedback to the existence of PORWs.

***Summary*** - The existence of PORWs as periodic highly localized pulse chains that appear randomly is demonstrated experimentally in long single-mode optical fibers for a continuous wave SBS system with weak pump feedback. Each pulse consists of a single but slightly modulated cycle of the optical wave envelope. The observed PORWs are also fairly well reproduced by an analytical model based on the three-wave coupling equations for the SBS with weak pump feedback added, which leads to modulation of the pump and anti-Stokes waves. For large pump powers, PORWs are found to occur with higher probability, making them fairly reproducible.

The present results may be useful for interpreting observations of rogue wave behavior in other nonlinear media, such as oceans and condensed matter. For example, the testimony of Cdt. Frédéric-Moreau seems to indicate that the train of three very large localized waves with very steep sides ("the Glorious Three") observed by the French cruiser *Jeanne d'Arc* on February 4, 1963 were periodic ocean rogue waves, as they have similar properties as the PORWs discussed here [32]. The rarity of periodic ocean rogue waves may then be associated with the fact that the probability of natural phase feedback control in freely ocean waves is rather small.

*Acknowledgments* - The authors acknowledge the support of the National Natural Science Foundation of China under the Grant Nos. 60978006, 11205194, 11247007, and 11374262, as well as the Beijing Natural Science Foundation under the Grant No. 4122055 and the Open Fund of the State Key Laboratory of High Field Laser Physics at the Shanghai Institute of Optics and Fine Mechanics.